\documentclass[11pt]{article}
\usepackage[top = 2 cm, bottom = 2.5 cm, left = 2.5 cm, right = 2 cm]{geometry}
\usepackage{authblk, cite, color, amssymb, amsmath, enumitem}
\usepackage[hypertex, colorlinks = true, linkcolor = blue, citecolor = red]{hyperref}
\usepackage[dvipsnames]{xcolor}

\begin{document}

\title{General spherically symmetric solution of Cotton gravity}

\author[1,2]{Merab Gogberashvili \thanks{gogber@gmail.com}}
\affil[1]{Javakhishvili Tbilisi State University, 3 Chavchavadze Ave., Tbilisi 0179, Georgia}
\affil[2]{Andronikashvili Institute of Physics, 6 Tamarashvili St., Tbilisi 0177, Georgia}

\author[1]{Ani Girgvliani \thanks{Ani.Girgvliani504@ens.tsu.edu.ge}}

\maketitle

\begin{abstract}

\noindent
In this paper we present the general spherically symmetric static solution to the vacuum equations of Cotton gravity. The obtained metric solution reveals the presence of singularities at the photosphere of a spherical source, which probably obstruct the formation of the stellar Schwarzschild-radius black holes. The solution is characterized by two integration constants, whose values can be restricted by association with the Hubble horizon. We examine the diverse features of the solution, including the long-range modifications to Newton's force through the incorporation of the velocity-squared repulsive term to model the dark energy.



\end{abstract}


\section{Introduction}

The search for a comprehensive theory of gravity is a central pursuit in theoretical physics, aiming to explain the fundamental nature of spacetime and its interactions with matter. While General Relativity has proven remarkably successful in describing the gravitational dynamics of stars and planets \cite{Will:2014kxa}, it faces problems when applied to larger scales in explaining galaxy dynamics and the accelerated expansion of the universe (see, for example, the reviews \cite{Arbey:2021gdg, Weinberg:2013agg}). This had prompted researchers to explore alternative theories of gravity without introducing the dark components of the universe (see the review \cite{Joyce:2014kja} and references therein). Among these alternative gravity theories, the Cotton gravity has attracted considerable attention \cite{Harada:2021bte}, since a variety of new physical solutions was obtained that can explain some large-scale properties of our universe \cite{Harada:2022edl}.

Among these alternative theories, Cotton gravity has garnered significant attention \cite{Harada:2021bte}. This interest stems from the discovery of a variety of new physical solutions that may explain certain large-scale properties of the universe \cite{Harada:2022edl}. In this study, we introduce a novel spherically symmetric static solution to the vacuum equations of Cotton gravity. Importantly, these equations can be reformulated as effective Einstein equations with a modified energy-momentum tensor \cite{Mantica:2022flg, Mantica:2023ihx}. The resulting metric reveals the existence of singularities on the photosphere, suggesting that within the framework of Cotton gravity, black hole solutions may deviate significantly from the predictions of General Relativity. The spherically symmetric metric we derive depends on two integration constants, whose values we estimate by associating them with the Hubble horizon.


\section{Cotton gravity}

One approach to develop alternative model of gravity, without fundamentally revising existing theory, involves using the once-contracted differential Bianchi identity:
\begin{equation} \label{Bianchi}
\nabla_\alpha R^\alpha_{~\beta\nu\mu} = \nabla_\mu R_{\nu\beta} - \nabla_\nu R_{\mu\beta}~,
\end{equation}
where $\nabla_\mu$ denotes the covariant derivative associated with the Levi-Civita connection. By replacing the Ricci tensors at the right side using the standard Einstein equations,
\begin{equation} \label{Einstein}
R_{\mu\nu} = 8\pi G \left(T_{\mu\nu} - \frac 12 g_{\mu\nu}T\right)~,
\end{equation}
we obtain the third order to the metric tensor differential equations (first order with respect of the Riemann tensor). If additionally we express the Riemann tensor by the Weyl tensor, $W_{\alpha\beta\nu\mu}$, the condition \eqref{Bianchi} obtains the form of so-called quasi-Maxwellian equations of gravity:
\begin{equation} \label{Maxwell}
\nabla^\alpha W_{\alpha\mu\nu\sigma} = 4\pi G\, M_{\sigma\mu\nu} ~,
\end{equation}
where the gravitational 'current',
\begin{equation} \label{M}
M_{\sigma\mu\nu} = \nabla_\mu \left( T_{\sigma\nu} - \frac{1}{3}\,g_{\sigma\nu} T\right) - \nabla_\nu \left( T_{\sigma\mu} - \frac{1}{3}\,g_{\sigma\mu} T \right)~,
\end{equation}
is covariantly conserved quantity,
\begin{equation}
\nabla^\sigma M_{\sigma\mu\nu} = 0~.
\end{equation}
The quasi-Maxwellian equations of gravity \eqref{Maxwell} have been considered since a long time ago (see, for example, \cite{Eisenhart, Lichnerowicz, Jordan, Mitskevich}), though without any  real physical applications. Although they are obtained using standard Bianchi identities and Einstein equations, and hold for arbitrary solutions of Einstein’s equations, the model is not identical to General Relativity. In particular, in addition to \eqref{Einstein}, one can insert at the right side of \eqref{Bianchi} an arbitrary Codazzi term $\mathbf C_{\mu\nu}$, i.e. a tensor that satisfies the condition \cite{Lovelock}:
\begin{equation} \label{CODAZZI}
\nabla_\alpha \mathbf C_{\mu\nu} = \nabla_\mu \mathbf C_{\alpha\nu}~,
\end{equation}
which introduces additional flexibility, allowing for a broader range of solutions. Also, due to the third-order character of the equations \eqref{Maxwell}, initial and boundary problems should be modified and not all singular metric solutions to the Einstein equations, as well as singular coordinate transformations used for showing geodesic completeness, are admissible.

Among recent third order formalisms, the Cotton gravity has gained considerable attention \cite{Harada:2021bte, Harada:2022edl}. This model introduces modifications to the field equations by employing the Cotton tensor, $C_{\mu\nu\sigma}$, as a replacement for the Einstein tensor and incorporates the energy-momentum vorticity tensor \eqref{M} as a source:
\begin{equation} \label{HEq}
C^\sigma_{~\nu\mu} = 8\pi G M^\sigma_{~\nu\mu}~.
\end{equation}
The condition $g^{\nu\mu} C^\sigma_{~\nu\mu} = 0$ ensures that the standard energy-momentum conservation, $\nabla_\mu T^{\mu\nu} = 0$, is preserved. The Cotton tensor can be expressed using the Weyl tensor,
\begin{equation} \label{Cotton_tensor}
C_{\mu\nu\sigma} = 2 \nabla^\alpha W_{\alpha\mu\nu\sigma} ~,
\end{equation}
which relates the equations of Cotton gravity \eqref{HEq} to the quasi-Maxwellian equations \eqref{Maxwell}.

The third-order nature of the Cotton gravity equations \eqref{HEq} is primarily due to the introduction of Codazzi tensors \eqref{CODAZZI}, with the combination
\begin{equation} \label{}
\mathbf C_{\mu\nu} = R_{\mu\nu} - T_{\mu\nu} - \frac 16 \, g_{\mu\nu} (R - 2T)~,
\end{equation}
which serves as a specific example of such a tensor. As a result, the equations \eqref{HEq} can be reformulated as effective Einstein equations with a modified energy-momentum tensor \cite{Mantica:2022flg, Mantica:2023ihx}:
\begin{equation} \label{Eff-Einstein}
R_{\mu\nu} - \frac{1}{2}g_{\mu\nu} R = 8\pi G \, T_{\mu\nu} + \mathbf C_{\mu\nu} - g_{\mu\nu} \mathbf C~,
\end{equation}
where $\mathbf C = g^{\mu\nu}\mathbf C_{\mu\nu}$.

The vacuum equations of Cotton gravity,
\begin{equation} \label{Vacuum}
C^\sigma_{~\nu\mu} = 0~,
\end{equation}
are obtained when the generalized gravitational current tensor \eqref{M} vanishes, indicating that the energy-momentum tensor itself is a Codazzi tensor,
\begin{equation}
\nabla_\mu T^\sigma_{~\nu} = \nabla_\nu T^\sigma_{~\mu}~.
\end{equation}
This condition implies that the Ricci tensor is also a Codazzi tensor and is equivalent to Yang's pure space equations \cite{Yang:1974kj},
\begin{equation}
\nabla_\mu R^\sigma_{~\nu} = \nabla_\nu R^\sigma_{~\mu}~,
\end{equation}
which are also a third order generalization of Einstein's equations (requiring the scalar curvature $R$ to be constant \cite{Guilfoyle:1998wj}).

Problem with Yang's model, just like with many other alternative theories of gravity, is the violation of Birkhoff's theorem -- spherical symmetry does not automatically require solutions to be static \cite{Pavelle:1976nf}, which causes serious deviations from the observed characteristics of gravitational field already on the scale of the solar system.
The question of whether time-dependent solutions exist in Cotton gravity and whether Birkhoff's theorem generalizes to this theory is an open and challenging problem. The non-linearity and complexity of the relevant equations make this a difficult question to address definitively.

In a special case where $T_{\mu\nu}$ is not a Codazzi tensor, and $\mathbf C_{\mu\nu} = 0$, Cotton gravity equations \eqref{Eff-Einstein} yield the standard Einstein equations with a source. On the other hand, $\mathbf C_{\mu\nu} = \Lambda g_{\mu\nu}$ corresponds to the Einstein equations with the cosmological term $\Lambda$. Notably, in the context of vacuum Cotton gravity \eqref{Vacuum} there exists some residual Riemann curvature (long-range gravitational induction in quasi-Maxwellian interpretation \cite{Lichnerowicz}), which is absent for vacuum Einstein equations.

Overall, the reason why Cotton gravity may deserve attention, lies in its variety of new solutions and interesting features that come with it, while retaining the core advantages of General Relativity. Below we shall present some exact solutions of vacuum Cotton equations \eqref{Vacuum}, which also enfold the case with constant Ricci scalar that in General Relativity corresponds to the constant energy distribution. This means that in the Cotton model cosmological term appears as a constant of integration, which provides a notable theoretical advantage over General Relativity in certain contexts.


\subsection{Axially symmetric solution}

At first note that general axially symmetric static solution to the vacuum Cotton equations \eqref{Vacuum} is done by the usual Kerr-de Sitter metric \cite{Carter},
\begin{equation} \label{kdsstandard}
\begin{split}
ds^2_{\rm KdS} &= \frac{\Delta - \left(1 + \Lambda a^2\cos^2\theta \right)a^2\sin^2\theta}{\rho^2 \left(1 + \Lambda a^2\right)^2}\, dt^2 + \frac{2a \sin^2\theta \left[\left(1 + \Lambda a^2\cos^2\theta \right)\left(\mathfrak{r}^2 + a^2\right) - \Delta \right]}{\rho^2 \left(1 + \Lambda a^2\right)^2}\, dtd\phi \,- \\
     &- \frac{\rho^2}{\Delta}\,d\mathfrak{r}^{2} - \frac{\rho^2}{1 + \Lambda a^2 \cos^2\theta}\,d\theta^2 - \frac{\sin^{2}\theta \left[\left(1 + \Lambda a^2 \cos^2\theta \right) \left(\mathfrak{r}^2 + a^2 \right)^2 - \Delta a^2 \sin^2 \theta \right]}{\rho^2 \left(1 + \Lambda a^2 \right)^2}\,d\phi^2 ,
\end{split}
\end{equation}
where $\mathfrak{r}$ is the Boyer-Lindquist radial coordinate \cite{Boyer:1966qh}, and
\begin{equation} \label{dr}
\begin{split}
\rho^2 = \mathfrak{r}^2 + a^2 \cos^2\theta ~, \qquad \Delta = \mathfrak{r}^2 - 2Gm\mathfrak{r} + a^2 - \Lambda \mathfrak{r}^2 \left(\mathfrak{r}^2 + a^2 \right)~,\qquad a = \frac Jm ~.
\end{split}
\end{equation}
The metric \eqref{kdsstandard} can be employed to model the gravitational field around rotating objects with the mass $m$ and the angular momentum $J$ in the presence of a cosmological constant $\Lambda$. Within General Relativity, it is challenging to find a rotating interior solution that can be matched to the Kerr exterior or any asymptotically flat vacuum exterior solution. We hope that Cotton gravity may provide a satisfactory solution to this problem.


\subsection{Spherically symmetric case}

It was found that the vacuum Cotton gravity equations \eqref{Vacuum} admit a Schwarzschild-type static spherically symmetric solution \cite{Harada:2021bte}:
\begin{equation} \label{Schwarz}
ds^2_{\rm Sch} = (1 - A) dt^2 - \frac{1}{(1 - A)} dr^2 - r^2 (d\theta^2 + \sin^2\theta d\phi^2)~,
\end{equation}
with
\begin{equation} \label{Harada}
A(r) = \frac{2Gm}{r} + \gamma r - \Lambda r^2~,
\end{equation}
where $\gamma$ and $\Lambda$ are integration constants. It is worth mentioning that the same function \eqref{Harada} appears in the general spherically symmetric solutions of conformal Weyl gravity \cite{Mannheim:1988dj}, as well as in the model for gravity at large distances \cite{Grumiller:2010bz}. The novel feature of the solution \eqref{Harada} is the appearance of the linear term $\gamma r$, which is absent in Einstein's gravity. In conventional approaches, similar terms are introduced to account for the effects attributed to dark matter halos. The solution \eqref{Harada} (with $\Lambda = 0$)  has been successfully used to model the rotation curves of galaxies, which are a manifestation of dark matter effects \cite{Harada:2022edl}.

In order to estimate the integration constants $\gamma$ and $\Lambda$ in \eqref{Harada}, we note that the natural geometric interpretation of the $\gamma r$ term is the Rindler-type acceleration, as the line-element \eqref{Schwarz} reduces to the Rindler metric in the $(t-r)$ plane when $m = \Lambda = 0$ \cite{Wald}. The exterior region of a gravitating object may be viewed as a reference fluid, moving toward the spherical source with a constant radial velocity $|v_r| \sim \sqrt{2Gm/r}$. In this picture, $\gamma r$ term in \eqref{Harada} can be understood as a small radial acceleration of space.

Since $\gamma r$  and $\Lambda r^2$ terms increase with distance, the Hubble radius should serve as a natural cutoff for solutions like \eqref{Harada}. When $r$ becomes of the order of the radius of the visible Universe, the product $\Lambda r^2$ in \eqref{Harada} reaches the order of unity, consistent with the observed value of the cosmological constant $\Lambda \approx 10^{-123}$ \cite{SupernovaSearchTeam:1998fmf, SupernovaCosmologyProject:1998vns}. If we assume a similar behavior for the $\gamma r$ term, we can estimate $\gamma \approx 10^{-62} - 10^{-61}$, which is approximately the scale at which the anomalous MOND ($\sim 10^{-62}$ \cite{Milgrom:1983ca}) and Pioneer ($\sim 10^{-61}$ \cite{Anderson:1998jd}) accelerations are observed. Therefore, it seems that in Cotton gravity the values of the integration constants $\gamma$ and $\Lambda$ are connected to the Hubble radius, which indicates the range of validity of the solution \eqref{Harada}.


\section{General spherically symmetric vacuum solution}

The simple spherically symmetric {\it ansatz} \eqref{Schwarz}, with $g_{tt}(r) = - 1/g_{rr}(r)$, is very restrictive for Cotton vacuum equations \eqref{Vacuum}. The general static spherically symmetric metric in Schwarzschild's gauge can be written as follows:
\begin{equation} \label{Anisotropic}
ds^2 = [1 - A(r)]\, dt^2 - \frac{1}{[1 - B(r)]}\, dr^2 - r^2 \left(d\theta^2 + \sin^2\theta d\phi^2 \right)~.
\end{equation}
The general solution to the vacuum Cotton equations \eqref{Vacuum} for the first metric function $A(r)$ is done by \eqref{Harada} (see the recent analyzes in \cite{Barnes:2023uru}). So, for simplicity, let us assume that $A(r)$ is expressed by \eqref{Harada} and display the solution to the vacuum Cotton equations \eqref{Vacuum} for the second unknown function $B(r)$ only:
\begin{equation} \label{B(r)}
\begin{split}
B (r) &= \frac {r^4 \left(\Lambda  r^3 - \gamma  r^2 - 2Gm + r \right) C_2 - \left( 3\gamma r^2 + 9Gm - 4r \right) \left( \Lambda r^3 - \gamma r^2 - 2Gm + r \right) C_1}{r \left( 2r - \gamma r^2 - 6Gm \right)^2} +\\
&+ \frac { \left( 51\gamma ^2\Lambda  + 68\Lambda ^2 + 60\gamma \Lambda ^2Gm \right)r^7 - \left( 51\gamma ^3 + 60\Lambda Gm\gamma ^2 \right) r^6}{17 r \left( 2r - \gamma r^2 - 6Gm \right)^2} - \\
&- \frac{\left( 120\Lambda G^2m^2\gamma  + 42\gamma ^2Gm - 80\Lambda Gm - 68\gamma  \right) r^4}{17 r \left( 2r - \gamma r^2 - 6Gm  \right)^2} -\\
&-\frac{ \left( 180\Lambda G^2m^2 + 208\gamma Gm \right) r^3 + \left( 56Gm - 300\gamma G^2m^2  \right) r^2 - 360G^3m^3}{17 r \left( 2r - \gamma r^2 - 6Gm \right)^2}~,
\end{split}
\end{equation}
where $C_1$ and $C_2$ are integration constants.

Note that the vacuum Cotton gravity equations \eqref{Vacuum} in the frame \eqref{Anisotropic} correspond to the Einstein equations with some exotic anisotropic fluid \cite{Mantica:2023ihx}:
\begin{equation} \label{T}
T_{\mu\nu} = \left(\varepsilon + p_r\right) u_\nu u_\mu - g_{\mu\nu} p_\bot + \left(p_r - p_\bot\right)\mathfrak{a}_\mu \mathfrak{a}_\nu ~.
\end{equation}
Here energy density, radial and tangential pressures have the expressions:
\begin{equation} \label{}
\begin{split}
\varepsilon &= (1-B) \left[\frac {1}{r^2} - \frac {B'}{r(1-B)}\right] + \frac {1}{r^2}~, \\
p_r &= (1-B) \left[\frac {1}{r^2} - \frac {A'}{r(1-A)}\right] - \frac {1}{r^2}~, \\
p_\bot &= (1-B) \left[-\frac {A''}{2(1-A)} + \frac {A'^2}{4(1-A)^2} - \frac {A'}{2r(1-A)} + \frac {A'B'}{4(1-A)(1-B)}- \frac {B'}{2r(1-B)}\right] ~, \\
\end{split}
\end{equation}
where primes denote radial derivatives and the functions $A(r)$ and $B(r)$ are done in \eqref{Harada} and \eqref{B(r)}, respectively. The four-velocity of the fluid, $u_\mu$, is a time-like vector ($u^\nu u_\nu = 1$) and $\mathfrak{a}^\nu$ is the unit space-like vector ($\mathfrak{a}^\nu \mathfrak{a}_\nu = -1$) in the radial direction ($u^\nu \mathfrak{a}_\nu = 0$, proportional to the four-acceleration), which on the background \eqref{Anisotropic} can be expressed as
\begin{equation}
u^\nu = \frac {1}{\sqrt {1-A}}\,\delta^\nu_t ~, \qquad \mathfrak{a}^\nu = \sqrt {1-B}\,\delta^\nu_r~.
\end{equation}

The solution \eqref{B(r)} looks complicated and let us analyze some particular, physically interesting cases.

When the metric function \eqref{Harada} is zero, i.e. $m = \gamma = \Lambda = 0$, the second metric function \eqref{B(r)} obtains the form:
\begin{equation} \label{A=0}
B_{\rm 0}(r) = \frac {C_2}{4} r^2 + \frac {C_1}{r}~.
\end{equation}
If the integration constant $C_1$ is also zero, the metric \eqref{Anisotropic} will represent Einstein’s static universe, which is a vacuum solution in Cotton theory!

For the pure Schwarzschild case with $\gamma = \Lambda = 0$, the metric functions
\begin{equation} \label{A,B-Sch}
\begin{split}
A_{\rm Sch}(r) &= \frac {2Gm}{r}~, \\
B_{\rm Sch}(r) &= \frac {r^4 (r - 2Gm) C_2 + (4r - 9Gm)(r - 2Gm) C_1}{4r(r - 3Gm)^2} + \frac {2Gm (45G^2m^2 -7r^2)}{17r(r - 3Gm)^2}~,
\end{split}
\end{equation}
reduce to the familiar Schwarzschild expressions, $A_{\rm Sch}(r) = B_{\rm Sch}(r) = 2Gm/r$, only if we set $C_1 = 48Gm/17$ and $C_2 = 0$. Important difference of the function $B_{\rm Sch}$ with the standard Schwarzschild vacuum solution is the appearance of the function $(3Gm - r)$ in the denominator. This factor also presents in all geometric invariants, e.g. Ricci and Kretschmann scalars. This means that the solution \eqref{A,B-Sch} leads to unavoidable singularity already on the photon sphere, $r = 3Gm$ \cite{Claudel:2000yi}, which prevents geodesics to be continued up to the central singularity at $r=0$ and can be addressed to the black holes information paradox.

In general, in addition to the vacuum spherically symmetric solution one needs to find an interior metric (describing configurations of stellar objects) and match it with the exterior solution \eqref{A,B-Sch}. Admitted model for the structure of relativistic stellar objects is unknown and very few of all known exact spherically symmetric solutions of the Einstein equations satisfy to conditions for the physically reasonable metric functions close to the junction surface, making it difficult to match internal and external solutions \cite{Del-Lake}. In many models the obtained junction radii allow matching of the internal and external Schwarzschild's solutions even inside the photon sphere $r = 3Gm$ \cite{Gogberashvili:2018tzc}. However, the photon sphere is the lower bound for any stable orbit and it seems that such a static object, similar to a black hole, cannot be realistic. More general spherically symmetric Cotton solution \eqref{A,B-Sch} can address this problem, although in physically relevant cases metrics can be continuously matched only outside the photon sphere of the object.

When considering the case without ordinary matter ($m = 0$), the metric functions in \eqref{Anisotropic} obtain the form:
\begin{equation} \label{A,B-D}
\begin{split}
A_{\rm Dark}(r) &= \gamma r - \Lambda r^2\,, \\
B_{\rm Dark}(r) &= \frac { r^3 \left( \Lambda r^2 - \gamma r + 1 \right) C_2 + \left( \Lambda r^2 - \gamma r + 1 \right) \left( 4 - 3\gamma r \right) C_1}{r\left( 2 - \gamma r \right)^2} +\\
&+ \frac {r \left( 3\gamma^2\Lambda r^3 - 3\gamma^3r^2 + 4\Lambda^2 r^3 + 4\gamma \right) }{ \left( 2 - \gamma r \right)^2}\,.
\end{split}
\end{equation}
We see that in the expression of $B_{\rm Dark}(r)$ there are cross terms $\gamma \Lambda$, indicating possible interactions between dark matter and dark energy in Cotton gravity.

Another interesting scenario is when the cosmological constant-like term is absent ($\Lambda = 0$) in \eqref{Anisotropic}. In this case, which was considered to address the problem of galaxy rotation curves in the context of Cotton gravity \cite{Harada:2021bte, Harada:2022edl}, the metric functions are given by
\begin{equation} \label{A,B-Harada}
\begin{split}
A_{\rm Harada}(r) &= \frac {2Gm}{r} + \gamma r~,\\
B_{\rm Harada}(r) &= \frac { r^4\left( 2Gm + \gamma r^2 - r \right)  C_2 + \left( 2Gm + \gamma r^2 - r \right)  \left( 9Gm + 3\gamma r^2 - 4r \right) C_1}{r \left( 6Gm + \gamma r^2 - 2r \right)^2 } \,- \\
&- \frac {51\gamma^3 r^6 - (42\gamma^2 Gm - 68\gamma) r^4 + 208 \gamma Gm r^3 + (56 Gm - 300 \gamma G^2m^2) r^2 - 360 G^3 m^3}{ 17r \left( 6Gm + \gamma r^2 - 2r \right)^2}~.
\end{split}
\end{equation}
At large distances, where $2Gm/r \to 0$ and $\gamma r \lesssim 1$, if we put $C_1 = 48Gm/17$ as for the pure Schwarzschild case, the second metric function from \eqref{A,B-Harada} behaves as:
\begin{equation} \label{B-H}
B_{\rm Harada}(r) \approx  \frac {2Gm}{r} + \gamma r + \frac {C_2}{4} r^2~,
\end{equation}
where $C_2$ is an arbitrary constant. This function differs from the solution considered in \cite{Harada:2021bte, Harada:2022edl} with the last term.


\section{Velocity squared force}

Let us now consider the weak gravity approximation of the non-vacuum Cotton equations \eqref{HEq} for the pure Schwarzschild-type metric \eqref{A,B-Sch},
\begin{equation} \label{metric}
ds^2 \approx [1 - A_{\rm Sch}  (r)]dt^2 - [1 + B_{\rm Sch}  (r)]dr^2 - r^2 (d\theta^2 + \sin^2\theta d\phi^2)~.
\end{equation}
The non-zero component of Cotton gravity equations is \cite{Harada:2021bte, Harada:2022edl}:
\begin{equation}
\nabla_r \left(R_{00} + \frac{1}{6} R \right) = -\frac{16\pi G}{3} \nabla_r \rho
\end{equation}
($\rho$ represents energy density of a non-relativistic source), or equivalently
\begin{equation} \label{eq:FE3}
R_{00} + \frac{1}{6} R = -\frac{16\pi G}{3} \rho~,
\end{equation}
where the integration constant was chosen zero. Using the metric \eqref{metric}, for the nonzero component of the Ricci tensor we obtain:
\begin{equation}
R_{00} \approx \frac{1}{2}\nabla_r^2 A_{\rm Sch}  = \frac{1}{2r^2} \left(r^2 A_{\rm Sch} ^\prime \right)^\prime = \frac{1}{2}A_{\rm Sch} ^{\prime\prime} + \frac{A_{\rm Sch} ^\prime}{r}~,
\end{equation}
where the primes denote derivatives with respect to the radial coordinate $r$.

Assuming Einstein's equations are satisfied and considering the metric \eqref{metric} with $A_{\rm Sch} (r) = B_{\rm Sch} (r)$, the expression \eqref{eq:FE3} reduces to the usual Poisson equation. However, for the general Cotton case \eqref{metric}, the Ricci scalar is given by
\begin{equation} \label{eq:Ricciscalar_linear}
R \approx - B_{\rm Sch} ^{\prime\prime} - \frac{4B_{\rm Sch} ^\prime}{r} - \frac{2B_{\rm Sch} }{r^2} + \nabla_r^2 (A_{\rm Sch} - B_{\rm Sch} )~,
\end{equation}
and the field equation \eqref{eq:FE3} obtains the form:
\begin{equation} \label{eq:EFE}
A_{\rm Sch} ^{\prime \prime} + \frac {2A_{\rm Sch} ^\prime}{r} - \frac{1}{2} \left(B_{\rm Sch} ^{\prime\prime} + \frac{3B_{\rm Sch} ^\prime}{r} + \frac{B_{\rm Sch} }{r^2} \right) = - 8\pi G \rho~.
\end{equation}
Assuming that the potentials $A_{\rm Sch} (r)$ and $B_{\rm Sch} (r)$ contain identical Newtonian terms, the solution to \eqref{eq:EFE} is given by:
\begin{equation} \label{solution}
A_{\rm Sch}  = \frac{2Gm}{r}~, \qquad B_{\rm Sch} = \frac{2Gm}{r} + \frac {C_2}{4} r^2~, \qquad 8\pi G\rho = \frac {9C_2}{4}~.
\end{equation}
It is important that in \eqref{solution}, the cosmological constant-like term,
\begin{equation}
\frac {C_2}{4} r^2 = \frac {8\pi G \rho}{9} r^2~,
\end{equation}
enters only the expression of the second metric function $B_{\rm Sch}$. Then, one needs to consider the velocity dependent term in the geodesic equations. Indeed, it is known that under the assumptions of a static spacetime and small curvature, the geodesic equation for a test mass can be approximated as \cite{Wald}:
\begin{equation} \label{Geodesic}
\frac{d^2x^i}{dt^2} \approx - \frac {dx^\mu}{dt} \frac {dx^\nu}{dt} \Gamma^i_{~\mu\nu} \approx - \Gamma^i_{~00} - \Gamma^i_{~jk} v^jv^k~,
\end{equation}
where $v^j \ll 1$ denotes 3-velocity, which is typically ignored in Newton's approximation when $g_{00}' \approx g_{rr}'$. However, in the spherically symmetric case of Cotton gravity \eqref{solution}, for the radial acceleration we obtain:
\begin{equation} \label{a_r}
a_r \approx \frac 12 \left( g_{00}' - g_{rr}' v^2_r\right) \approx - \frac {Gm}{r^2}  + \frac {32\pi G \rho}{9}\, r \,v^2_r~.
\end{equation}
The interesting feature of this solution is the appearance of velocity-squared repulsive long-range corrections to Newton's gravitational force. In Cotton gravity, the extra long-range curvature can be interpreted as an effective perfect fluid filling the universe \eqref{T}, giving rise to this additional force. The appearance of such correction is natural, as velocity-squared forces are known to exist in other physical systems, such as drag forces on objects moving at high speeds relative to a surrounding fluid \cite{Knight}.

Velocity-dependent forces, like the Coriolis and Lorentz forces, are well-known, but velocity-squared forces are relatively rare. The concept of a gravitational force proportional to the square of the velocity was initially proposed by Schr\"odinger \cite{Schrodinger}, and more recently, a repulsive force proportional to the squared velocity dispersion of a structure was derived by contracting the relativistic generalization of angular momentum \cite{Hansen:2021fgb}.

Interestingly, it has been suggested that a universe without dark energy, but instead featuring a repulsive force proportional to the velocity squared, could exhibit an accelerated expansion that closely mimics the effects of a cosmological constant \cite{Loeve:2021ozs}. This idea provides an alternative explanation for the observed accelerated expansion of the universe, highlighting the potential implications of velocity-squared forces in gravitational theories. The presence of velocity-squared forces, in addition to the cosmological constant, highlights the richness of Cotton gravity and expands the range of possibilities for cosmological studies.


\section{Conclusions}

The paper discussed different solutions in the context of Cotton gravity. We have obtained the general spherically symmetric solution to the vacuum equations of the model. The derived metric exhibits intriguing properties, notably the presence of singularities on the photosphere, which seems to prevent formation of the Schwarzschild-radius stellar black holes. This suggests that within the framework of Cotton gravity, black hole solutions may exhibit significant deviations from the predictions of General Relativity. Obtained spherically symmetric metric depends on two integration constants, which have been associated with the Hubble horizon to estimate their values. It was noted that Cotton gravity may also offer a solution for finding realistic interior metrics.

The velocity squared force within the Cotton gravity framework is examined. Geodesic equations lead to the inclusion of the repulsive velocity-squared term in the gravitational acceleration, which results in long-range modifications to Newton's gravitational potential, providing an interesting departure from the standard gravitational theory. The appearance of these additional long-range curvature effects is attributed to the presence of an effective perfect fluid in the Universe. The concept of velocity-squared forces in gravitational theories is highlighted, drawing connections to other physical systems and alternative explanations for the observed accelerated expansion of the universe.

Our findings underscore the significance of the Cotton model as an alternative gravitational theory. The general spherically symmetric solution in Cotton gravity unveils a rich interplay between geometry, gravity, and the dynamics of the universe and prompts us to reconsider our understanding of gravity at both small and large scales. Future investigations in this area hold the potential to deepen our knowledge of fundamental physics and cosmology, offering new insights into the mysterious components of the universe and the dynamics of gravitation in the realm of the Cotton model.


\section*{Declarations:}

\subsection*{Ethical approval}
Not applicable.

\subsection*{Conflict of interests}
Both authors declare that they have no conflicts of interest.

\subsection*{Authors' contributions}
M.G. developed the theoretical formalism and performed the analytic calculations. Both authors, M.G and A.G., contributed to the final version of the manuscript. M.G. supervised the project.

\subsection*{Funding}
Not applicable.

\subsection*{Availability of data and materials}
There are no data associated with this article.


\end{document}